\documentclass[aps,amsmath,amssymb,showpacs,showkeys]{revtex4}
\usepackage{graphicx}

\begin{document}

\title{ Evaluation of the electron-TO-phonon interaction \\
in polar crystals from experimental data }

\author{Aleksandr Pishtshev}
\email[]{E-mail: ap@eeter.fi.tartu.ee}
\affiliation{Institute of Physics, University of Tartu, Riia 142, 51014 Tartu, Estonia}

\begin{abstract}
\noindent
The aim of this study was to determine the strengths of the coupling of electrons
with the polar long-wavelength transverse optical (TO) vibrations
from infrared spectroscopy data.
This determination is made by means of a simple relationship 
between the electron-TO-phonon interaction constant and material parameters,
based on a parametrization of the electron-TO-phonon coupling
in terms of the long-range dipole-dipole interaction.
The combination of experimental data employed here allowed us to calculate directly
the relevant constants for a number of representative polar insulators and
to show that in ferroelectrics the interband electron-TO-phonon interaction
at the $\Gamma$ point is essentially strong.
In these calculations, infrared spectroscopy methods proved to be an effective tool
for study of the main properties of electron interaction
with polar long-wavelength TO phonons.
\end{abstract}


\maketitle

%
\section{Introduction}
In solid state science, there are a number of techniques
for probing detailed characteristics related to the structural and
dynamical properties of materials, which are therefore capable of numerical prediction
of electron-phonon coupling constants from experimental data.
In particular, methods of vibrational spectroscopy have been shown
to yield reliable estimates of several fundamental microscopic parameters,
including relevant constants of electron-phonon interactions
(e.g.,~\cite{Peticolas,Graja,Tanner,Ferrari,Girlando} and references therein).

Two common properties of the lattice dynamics of a polar crystal are 
(i) the existence of dipole moments associated with
optical vibrational modes and
(ii) the difference in long-range fields given by longitudinal and
transverse optical modes at the zone center~\cite{BornM}.
As shown in our previous study~\cite{Pishtshev},
in polar materials, the coupling of electrons with the polar
long-wavelength transverse optical (TO) vibrations is essentially influenced by the 
features of the Coulomb interaction between electrons and the lattice ions. 
In turn, these features are directly related to the presence of the dipole moments
of the TO vibrational modes.
At the same time, dependencies of the dipole moments on the ionic degrees of freedom
give rise to infrared (IR) spectra~\cite{Robinson,Stuart},
which makes IR spectroscopy an important tool
for investigating the low-frequency dynamics of the IR-active 
TO modes in polar materials (e.g.,~\cite{Kamba}).

The characteristics of the IR spectra can be understood
in terms of effective charges belonging to the TO vibrational modes.
The relative values of these charges are directly related to the strength
of the corresponding long-range electron-TO-phonon couplings~\cite{Pishtshev}.
By combining these features, we see that information on the intensities
of IR-active vibrations provided by IR spectroscopy measurements
can be successfully used for quantitative modeling of the interactions
of electrons with zone-center TO phonon modes.
Thus, the analysis of the IR spectra data in terms of the dielectric function
can serve as an alternative to ab-initio methods to obtain from an experiment
independent estimates of the electron-TO-phonon (el-TO-ph) coupling constants
in a given material.

To the best of our knowledge, with only a few exceptions,
no direct estimates of the el-TO-ph interaction constants were previously available
to provide numerical values for most polar materials, especially for ferroelectric systems
in which the strength of the el-TO-ph interaction should be especially strong~\cite{Pishtshev}.
In particular,
even for several well-studied compounds, such as $\rm BaTiO_3$, $\rm SrTiO_3$,
and $\rm SnTe$, there was little information on how different or similar the relevant
constants of the electron-TO phonon coupling are.
In this work, we have performed a numerical investigation
of the el-TO-ph interaction constants for a number of representative polar compounds
using the theoretical scheme proposed in~\cite{Pishtshev}.
To obtain a more detailed picture, we present a comparative analysis of the results
of our calculations. As expected~\cite{Pishtshev}, very large values of the interband el-TO-ph
interaction constants were found to be a general feature of ferroelectric materials.
%
\section{Theoretical details}
Contributions of the IR active optical phonons to the dielectric function
${\epsilon}({\omega})$ of the far-IR spectral range, which are the subject
of IR spectroscopy measurements and analysis, are usually modeled by a classical damped
harmonic oscillator model~\cite{Cowley2,Kamba}:
\begin{equation}\label{dm-eq1} 
{\epsilon}({\omega}) \,=\, {\epsilon}_{\infty} \,+\,
{\sum_{j}}\, 
\frac {S(j)} {\, \Omega^{2}_{j}{\,-\,}{\omega^{2}} {\,-\,} {i{\gamma}_j{\omega}} \,}
\end{equation}
with ${\Omega}_j$, ${\gamma}_j$ and $S(j)$ the $j$-th zone-centre TO vibrational
mode frequency, damping and dipole oscillator strength, respectively;
${\epsilon}_{\infty}$ is the core (electronic high-frequency) contribution
to the dielectric function.

In our previous work~\cite{Pishtshev}, within a first-principles methodology,
we predicted theoretically the significant increase of the constants
of the el-TO-ph coupling in ferroelectric materials, showed how the el-TO-ph interaction
constants might be dependent on macroscopic material parameters, and obtained
analytical equations allowing us to estimate el-TO-ph interaction strengths
in a wide range of polar dielectrics. In the limit of a zero phonon wave-vector,
the el-TO-ph interaction represents the long-range electron-lattice dynamic hybridization
of the electronic bands of opposite parities, which takes into account the relevant
$s$-, $p$- and $d$-channels (e.g.,~\cite{Kristoffel1,Kristoffel2,Bersuker1,Bersuker2}
and references therein):
\begin{equation}\label{hamiltonian_el-TO-ph}
H_{el-ph} =
{N^{-1/2}} {\sum_{\sigma \neq \sigma^{'} }}\,{\sum_{\bf k}}\,
g^{j}_{{\sigma}{\sigma^{'}}}({\bf k})\,
a^+_{\sigma{\bf k}}a_{\sigma^{'}{\bf k}} \, y_{0j} \,.
\end{equation}
Here the matrix elements of the el-TO-ph interaction at the $\Gamma$ point,
$g^{j}_{{\sigma}{\sigma^{'}}}({\bf k})\,$, serve as the central parameters of
the effective Hamiltonian~\eqref{hamiltonian_el-TO-ph}, which characterize
the interband scatterings of the bond electrons due to the zone-center TO phonons;
$a^+$ ($a$) are the creation (annihilation) operators
for electronic states $|\,{\sigma}{\bf k}\,${\textgreater}
of the valence and conduction bands ($\sigma, \sigma^{'}$) in an insulator,
$y_{0j}$ denotes the zero-momentum TO mode normal coordinates,
and $N$ is the number of unit cells.
For instance, as applied to a family of high polar crystals such as 
the ferroelectric $\rm ABO_{3}$ perovskite oxides,
hybridization~\eqref{hamiltonian_el-TO-ph} involves significant mixing
between the filled $\rm O$ $\rm 2p$ and the empty ${\rm d}^{0}$
(${\rm Ti}^{4+}$, ${\rm Nb}^{5+}$, ${\rm Zr}^{4+}$, ${\rm Ta}^{5+}$,
${\rm Mo}^{6+}$, ${\rm W}^{6+}$, etc.) electronic states
caused by the IR-active TO ${\rm F}_{1u}$ soft
vibrations~\cite{Kristoffel2,Konsin1,Konsin2A}.

For the last several decades, the Fr\"{o}hlich-type model (\ref{hamiltonian_el-TO-ph})
was a useful prototype for intensive studies of various properties of both typical insulating
perovskites~\cite{Kristoffel2,Konsin1,Ohnishi,Hidaka,Bussmann-Holder2,Girshberg1}
and the $\rm A^{IV}B^{VI}$ narrow-gap semiconductors and their
alloys~\cite{Kawamura,Murase,Konsin2,Sakai,Maksimenko}. It was recently demonstrated
to be applicable to give a theoretical description of the ferroelectricity found in 
$\rm BiFeO_{3}$-type multiferroics~\cite{Konsin3,Konsin4}.
In light of these studies and of the fundamental significance of electron-phonon coupling
in the understanding of various physical properties of polar compounds (e.g.,~\cite{Mechelen}),
the essential question then naturally arises of what method could be applied to estimate
(regardless of a model used) the numerical values of the el-TO-ph interaction constants
${g_{j}}$ in a polar material. As shown in Ref.~\cite{Pishtshev},
one particular advantage in the study of TO lattice vibrations is that 
one can associate the interaction between electrons
and the polar long-wavelength TO phonons with the long-range dipole-dipole interaction.
This link, which was not anticipated by previous theoretical considerations,
provides the following relationship
\begin{equation}\label{ephto0}
g_{0j}^{2} \,=\, \frac{1}{12} {{E}_g} \, M_{j} \, S(j) \, , \quad
%
%
{g}_{j}^{2} \,=\,
{\frac{3{g}_{0j}^{2}}{{\epsilon}_{\infty}+2}} \, .
\end{equation}
Here the first equation relates, for each zone-center TO vibration 
of the branch $j$ and of the reduced mass $M_j$,
the bare constant of the el-TO-ph coupling
at the $\Gamma$ point,
$g_{0j}$, with the macroscopic material constants, such as
$S(j)$, ${\epsilon}_{\infty}$, and a bond-gap energy ${E}_g\,$;
$g_{0j}^{2}$ represents the effective ${\bf k}$-independent
interband el-TO-ph interaction
defined by a proper average of the squared el-TO-ph matrix elements
over the given electronic states:
\begin{equation}
g_{0j}^{2}=
{{\sum_{\sigma^{'}{>}\,\sigma}} \,
\frac{1}{2N} 
{\sum_{\bf k}}} \,
{\vert \, g^{j}_{{\sigma}{\sigma^{'}}}({\bf k}) \vert}^2\,
\frac{{E}_g}{ \,\left|\,{E_{\sigma^{'}}({\bf k})-E_{\sigma}({\bf k})} \right|\, } \, ,
\end{equation}
where $E_{\sigma}({\bf k})$ are the energies
of the Bloch states of wavevector ${\bf k}$ in the ${\sigma}$th band. 
The second equation in~(\ref{ephto0})
defines the renormalized el-TO-ph coupling constant ${g}_{j}$
that accounts for local-fields effects-induced partial screening
of the bare el-TO-ph interaction.

With the help of Eqs.~(\ref{ephto0}), one can directly evaluate
the el-TO-ph coupling constants via the dipole oscillator strengths.
The dipole oscillator strengths can in turn be readily determined through 
IR spectroscopy data analysis. 
Therefore, employing the relevant information from the IR spectra
concerning the IR-active optical phonons and the dielectric function behavior
in the far-IR spectral range, and using Eqs.~(\ref{ephto0})
in combination with the experimental data for ${\epsilon}_{\infty}$ and ${E}_g$,
enables us to determine numerical values of the el-TO-ph coupling constants
for polar systems of interest.
%
\section{Results and discussion}
We performed numerical calculations with Eqs.~(\ref{ephto0}) for representative
polar compounds belonging to several classes of materials:
typical insulating perovskites, the $\rm A^{IV}B^{VI}$ narrow-gap semiconductors,
and various non-ferroelectric wide-gap binary compounds.
The evaluation of $g_{0j}$ and $g_{j}$ for these
compounds involved the knowledge of material parameters ($S(j)$, ${\epsilon}_{\infty}$, ${E}_g$)
whose experimental values were taken from the literature or, when necessary,
calculated from the corresponding experimental data.
Lists of all compounds considered
and the calculated el-TO-ph coupling constants
and material parameters used are given in Tables 1 and 2.
\begin{table}[pt]\label{table1}
\caption{The calculated values of the el-TO-ph interaction constants
provided for each polar TO mode of a cubic phase of
the ferroelectric $\rm ABO_{3}$ perovskite oxides
and $\rm A^{IV}B^{VI}$ narrow-gap semiconductors.
The material parameters used are also given.}
{\begin{tabular}{@{}l*{15}ccccccccc@{}} \toprule
Compound \,& \,${\epsilon}_{\infty}$\,& \,${E}_g$\,& TO & $$ & $M_{j}S(j)$ & Ref. & $g_{0j}$ & $g_{j}$ & $ {\kappa}_j$ & $K_0$ \\
&  & (eV) & ($j$) & & (eV/{\AA}$^{2}$) & & (eV/\AA) & (eV/\AA) &  & (eV/{\AA}$^{2}$) \\ \colrule
${\rm BaTiO_3}$ & $5.2$ & $3.3$ & $1$ & $$  & $194.3$ & \cite{Barker} & $7.3$ & $4.7$ & $2.41$ & $2.80$ \\
              &\cite{Zhong}&\cite{Peacock}& $2$ & $$ &  $15.1$ &  & $2.0$ & $1.3$ & $$ & $$ \\
              &       &       & $3$ & $$ &  $7.8$ &  & $1.5$ & $0.9$ & $$ & $$ \\ \colrule
%
${\rm SrTiO_3}$ & $5.2$ & $3.3$ & $1$ & $$  & $211.6$ & \cite{Barker} & $7.6$ & $4.9$ & $2.42$ & $3.04$ \\
              &\cite{Zhong,Han}&\cite{Robertson}& $2$ & $$ &  $18.8$ &  & $2.3$ & $1.5$ & $$ & $$ \\
              &       &       & $3$ & $$ &  $18.1$ &  & $2.2$ & $1.4$ & $$ & $$ \\ \colrule
%
${\rm KNbO_3}$ & $4.7$ & $3.1$ & $1$ & $$  & $246.4$ & \cite{Fontana} & $8.0$ & $5.3$ & $3.27$ & $2.81$ \\
              &\cite{Zhong}&\cite{Hayashi}& $2$ & $$ &  $24.3$ &  & $2.5$ & $1.7$ & $$ & $$ \\
              &       &      & $3$ & $$ &  $33.0$ &  & $2.9$ & $2.0$ & $$ & $$ \\ \colrule
%
${\rm KTaO_3}$ & $4.6^a$ & $3.6$ & $1$ & $$  & $211.0$ & \cite{Spitzer} & $8.0$ & $5.4$ & $2.81$ & $2.85$ \\
              &       &\cite{Shimizu}& $2$ & $$ &  $24.3$ &  & $2.7$ & $1.8$ & $$ & $$ \\
              &       &      & $3$ & $$ &  $28.4$ &  & $2.9$ & $2.0$ & $$ & $$ \\
\colrule 
%
\multicolumn{11}{l}{${\rm A^{IV}B^{VI}}$ narrow-gap semiconductors}\\
${\rm PbTe}$ & $32.8$ & $0.31$ & $1$ & $$  & $187.7$ & \cite{Razeghi,Kvyatkovskii} & $2.20$ & $0.65$ & $0.50$ & $2.68$ \\ 
%
${\rm PbSe}$ & $22.9$ & $0.28$ & $1$ & $$  & $92.5$ & \cite{Razeghi,Kvyatkovskii} & $1.47$ & $0.51$ & $0.29$ & $3.15$ \\ 
%
${\rm PbS}$ & $17.2$ & $0.42$ & $1$ & $$  & $71.0$ & \cite{Razeghi,Kvyatkovskii} & $1.58$ & $0.62$ & $0.27$ & $3.46$ \\
%
${\rm SnTe}$ & $37.0$ & $0.35$ & $1$ & $$  & $147.4$ & \cite{Kvyatkovskii,Saini} & $2.07$ & $0.58$ & $0.33$ & $2.85$ \\
\multicolumn{11}{l}{$^a$ estimated as in Ref.~\cite{Zhong}}\\
\botrule
\end{tabular} }
\end{table}
\begin{table}[pt]\label{table2}
\caption{The calculated values of the el-TO-ph
interaction constants in a number of cubic binary crystal systems.
The material parameters used are also given.}
{\begin{tabular}{@{}l*{15}cccccccc@{}} \toprule
%
Compound \,& ${\epsilon}_{\infty}$ \,& ${E}_g$ \,& $$ & $MS$ & Ref. & $g_{0}$ & $g$ & $ {\kappa}$ & $K_0$ \\
&  & (eV) & & (eV/{\AA}$^{2}$) & & (eV/\AA) & (eV/\AA) &  & (eV/{\AA}$^{2}$) \\ \colrule
\multicolumn{5}{l}{Oxides of the alkaline earth metals}\\
${\rm MgO}$ & $2.95$ & $7.2$ & $$  & $36.2$ & \cite{Sun,Stoneham} & $4.7$ & $3.6$ & $0.19$ & $9.67$ \\ 
%
${\rm CaO}$ & $3.38$ & $6.2$ & $$  & $33.6$ & \cite{Hofm,Stoneham} & $4.2$ & $3.1$ & $0.24$ & $6.50$ \\ 
%
${\rm SrO}$ & $3.28$ & $5.3$ & $$  & $29.4$ & \cite{Hofm,Stoneham} & $3.6$ & $2.7$ & $0.26$ & $5.26$ \\
%
${\rm BaO}$ & $3.61$ & $4.0$ & $$  & $31.0$ & \cite{Hofm,Stoneham} & $3.2$ & $2.4$ & $0.33$ & $4.26$ \\ \colrule
%
\multicolumn{5}{l}{Transition-metal magnetic monoxides}\\
${\rm MnO}$ & $4.85$ & $3.6$ & $$  & $52.8$ & \cite{Rudolf,Rodl} & $4.0$ & $2.6$ & $0.23$ & $8.24$ \\
%
${\rm NiO}$ & $5.61$ & $3.7$ & $$  & $45.5$ & \cite{Hofm,Rodl} & $3.7$ & $2.4$ & $0.15$ & $9.93$ \\ 
%
${\rm CoO}$ & $5.00$ & $2.8$ & $$  & $45.1$ & \cite{Rudolf2,Rodl} & $3.2$ & $2.1$ & $0.17$ & $9.36$ \\
%
${\rm FeO}$ & $5.38$ & $2.4$ & $$  & $43.5$ & \cite{Hofm,Rodl} & $2.9$ & $1.9$ & $0.16$ & $9.05$ \\ \colrule
%
\multicolumn{5}{l}{Alkali halides}\\
${\rm LiCl}$ & $2.70$ & $9.4$ & $$  & $7.0$ & \cite{Zinenko,Poole} & $2.3$ & $1.9$ & $0.07$ & $5.33$ \\
%
${\rm NaCl}$ & $2.33$ & $8.5$ & $$  & $4.9$ & \cite{Kvyatkovskii,Poole} & $1.9$ & $1.6$ & $0.07$ & $4.03$ \\
%
${\rm KCl}$ & $2.20$ & $8.4$ & $$  & $3.7$ & \cite{Stoneham,Poole} & $1.6$ & $1.4$ & $0.08$ & $2.90$ \\
%
${\rm RbCl}$ & $2.20$ & $8.2$ & $$  & $3.5$ & \cite{Stoneham,Poole} & $1.5$ & $1.3$ & $0.08$ & $2.54$ \\ 
%
${\rm LiF}$ & $1.90$ & $13.7$ & $$  & $11.0$ & \cite{Stoneham,Zinenko} & $3.5$ & $3.1$ & $0.06$ & $11.08$ \\ 
%
${\rm NaF}$ & $1.74$ & $11.5$ & $$  & $7.5$ & \cite{Stoneham,Zinenko} & $2.7$ & $2.4$ & $0.07$ & $7.24$ \\
%
${\rm KF}$ & $1.85$ & $10.8$ & $$  & $5.9$ & \cite{Stoneham,Zinenko} & $2.3$ & $2.0$ & $0.08$ & $4.73$ \\ 
\colrule
%
\multicolumn{5}{l}{Thallium and cesium halides}\\
${\rm TlCl}$ & $4.76$ & $3.2$ & $$  & $13.7$ & \cite{book1} & $1.9$ & $1.3$ & $0.16$ & $3.21$ \\
%
${\rm TlBr}$ & $5.34$ & $2.65$ & $$  & $12.6$ & \cite{book4} & $1.7$ & $1.1$ & $0.15$ & $2.89$ \\
%
${\rm CsCl}$ & $2.63$ & $8.3$ & $$  & $4.4$ & \cite{Kvyatkovskii,Poole} & $1.7$ & $1.4$ & $0.09$ & $2.58$ \\ 
\botrule
\end{tabular} }
\end{table}

Additionally, in both tables we presented the values of the dimensionless
coupling parameter~${\kappa}_{j}={g}_{j}^{2}/({E}_g K_0)$~calculated
for low-frequency TO modes. 
Here $K_0=4{\pi}e^{2}/v_{o}$,
and $v_{o}$ is the cell volume with respect to the formula unit.
From a physical point of view,  $ {\kappa}_{j}$ is represented
as a ratio of the electronic contribution ${g}_{j}^{2}/{E}_g$,
determined by the electronic structure and the electron-ion potential,
and some lattice force constant $K_0$. 
This specially introduced parameter is of considerable interest due to its direct ability
to supply a measure of the zone-center TO phonon softening arising from the interband
scatterings of the bond electrons.
As seen from Tables 1 and 2, only the wide-gap ferroelectric materials
are characterized by the high values of the parameter ${\kappa}$.
Since in the cubic binary crystal systems one can set $j=1$,
we also omitted the indices $j$ throughout Table 2.
Note that for the most part of binary compounds represented in this table,
we calculated the dipole oscillator strengths $S$
with the following correspondence:
$({\Omega}_{LO}^{2}-{\Omega}_{TO}^{2})\,{{\epsilon}_{\infty}}=S$
which relates the squared frequencies of the longitudinal and transverse
zone-center optical vibrations, ${\Omega}_{LO}^{2}$ and ${\Omega}_{TO}^{2}$,
respectively, with $S$.

Based on the comparative analysis of the results obtained and summarized
in both tables, we can draw several general conclusions and discuss the role
of the el-TO-ph coupling related to the ferroelectricity phenomenon:\\
i) A simple calculation recently proposed
in~\cite{Pishtshev} as a quantitative measure of
the el-TO-ph interaction is shown to be applicable for a wide number of
selected polar insulators and semiconductors, including representative 
ferroelectric compounds.\\ 
ii) Among the compounds considered, it was clearly demonstrated that
the el-TO-ph coupling strength is small
in alkali halides, intermediate in binary oxides,
both the alkaline earth and 3d-transition metals,
and, in contrast to the first two cases, significantly large in ferroelectric perovskites.
Notice also that the values of the el-TO-ph coupling constants ${g}_{j}$
tabulated in Table 1 for ${\rm A^{IV}B^{VI}}$ compounds
agree with the relevant values from semi-empirical estimates~\cite{Kristoffel1}.
The values of ${g}_{j}$ given in Table 1 for the wide-gap ferroelectrics
can be compared with the only known semi-empirical estimate
of Ref.~\cite{Kristoffel1} for $\rm BaTiO_3$, which, in turn, should be recast
by applying the accurate magnitude of the effective charge:
using the value $8.33$ estimated from the data of Table 1, one obtains
the corrected estimate $4.2$ eV/{\AA}, which is 
close to the value $4.7$ eV/{\AA}~of Table 1 for $\rm BaTiO_3$. \\
iii) There is a nontrivial one-to-one correspondence with the zone-center
TO vibrational mode effective charge $Z^{*}$, or equally with the Born effective charge.
It is seen that the charge $Z^{*}$ sets the scale of the el-TO-ph coupling;
for instance, in pure ionic wide-gap materials, where the Born effective charges
are very close to the nominal rigid-ion values, the actual el-TO-ph coupling strength 
is small.
To further quantify the interplay between these physical quantities,
we plot their interrelation in Fig.~1, which uniquely displays
the substantial growth of the el-TO-ph coupling with a rise in the effective charge:
$\kappa \propto \left|Z^{*}\right|^{2}$, just as predicted~\cite{Pishtshev}.
Due to the anomalously large Born effective charges, which occur in ferroelectric
compounds~\cite{Zhong,Ghosez}, a marked increase
in interband hybridization leads to a drastic enhancement of the interband
el-TO-ph scatterings.
In the limit of pure ionic wide-gap compounds, multiband channels
of electronic scatterings associated with zone-center TO phonons turn out
to be marginal, as the electronic polarizability effects are not high enough
to exhibit a noticeable el-TO-ph hybridization between different bands.
Thus, our calculations show that strong el-TO-ph coupling should be a general
feature of wide-gap compounds with high electronic polarizabilities.

As a matter of fact, the same results hold in the case of ${\rm A^{IV}B^{VI}}$ narrow-gap
semiconductor systems where, however, one should take into account two factors:
smallness of ${E}_g$ and sufficiently high range of ${\epsilon}_{\infty}\,$.
It is easy to see from Table 1 that the relevant values of the product $M_{j}S(j)$
for both wide-gap and narrow-gap ferroelectrics appears to be close to each other.
At the same time, in non-ferroelectrics, as seen from Table 2, these values are
significantly smaller. Moreover, for both groups of ferroelectrics,
the unscreened dimensionless el-TO-ph coupling parameter,
${\kappa}{_j} ({\epsilon}_{\infty}+2)/{3} \,$,
as seen from Table 3, turns out to be about the same order of magnitude.
Meanwhile, for the non-ferroelectric compounds considered, the values of
this parameter differ significantly and are much smaller
than in the ferroelectrics.
\begin{table}[pt]\label{table3}
\caption{The calculated values of the unscreened dimensionless el-TO-ph coupling parameter
for a selected number of polar compounds in a cubic phase.}
{\begin{tabular}{@{}l*{10}cccccccccccc@{}} \toprule
 \,& ${\rm KNbO_3}$ \,& ${\rm BaTiO_3}$ \,& ${\rm PbTe}$ \,& ${\rm SnTe}$ \,& ${\rm PbSe}$ \,& ${\rm PbS}$ \,&  ${\rm BaO}$ \,& ${\rm MnO}$ \,& ${\rm SrO}$ \,& ${\rm TlCl}$ \,& ${\rm CsCl}$ \,& ${\rm LiF}$\\
 \colrule
${\kappa}{_j} \frac {({\epsilon}_{\infty}+2)} {3}$
 & $7.3$ & $5.8$ & $5.8$ & $4.3$  & $2.4$ & $1.7$ & $0.61$ & $0.53$ & $0.46$ & $0.36$ & $0.14$ & $0.08$ \\
\botrule
\end{tabular} }
\end{table}
\section{Conclusion}
To summarize the results from this study, we refer to our previous work~\cite{Pishtshev},
where it was shown that the zone-center TO vibrational mode effective charge $Z^{*}$
can be considered to be 
the key macroscopic parameter of the el-TO-ph coupling strength. As is well-known,
the Born effective charges play a fundamental role in determining the lattice-dynamical
properties of insulating crystals and are a powerful tool for investigating the dielectric
and ferroelectric properties of materials. In particular, they measure the coupling of IR
radiation to the optic phonons. In materials where the el-TO-ph coupling is operative,
it can be verified by spectroscopy measurements of the IR-active TO mode.

In this paper, we pioneered the application of experimental data to estimate
the interband el-TO-ph interaction constants in a number of representative polar compounds.
We sustained theoretical arguments~\cite{Pishtshev} with numerical calculations
of the interband el-TO-ph interaction constants and the further comparative analysis
that the large interband el-TO-ph interaction is a special microscopic feature
of the ferroelectric materials. 
Thus, one can conclude that the presence of anomalously large Born's effective charges
is the key signature that the interband el-TO-ph coupling is essentially strong
in a ferroelectric material. In contrast, in non-ferroelectrics, the strength
of the el-TO-ph interaction is not necessarily high enough due to their lower polar nature.
This result corresponds to the main assumption of the vibronic
theory~\cite{Kristoffel1,Kristoffel2,Bersuker1,Bersuker2} regarding the existence of
sufficiently strong interband el-TO-ph coupling in displacive ferroelectrics.
The strength of the interband el-TO-ph interaction can therefore serve
as a direct indicator of the extent to which a crystal lattice is close
to a possible ferroelectric instability.
\section*{Acknowledgments}
The author would like to thank A. Sherman and P. Rubin for attention to this work.
The work was supported by the ETF grant No. 6918.
%
%
%

\newpage

{\large\bf Figure captions}\\
\vspace*{10pt}

{\bf Fig. 1}: Dependence of the square root of the dimensionless el-TO-ph coupling parameter
       on zone-centre TO vibrational mode effective charge. 
Triangles shown on the figure are the values of 
$\left|Z^{*}\right|$ and $\sqrt{\kappa}$
determined separately for each compound.
The values of $\left|Z^{*}\right|$
are calculated from the experimental data given in Tables 1 and 2.
The direct line is the resulting fit.

\newpage
\begin{figure*}[ht]
\begin{center}
\includegraphics[width=0.9\textwidth]{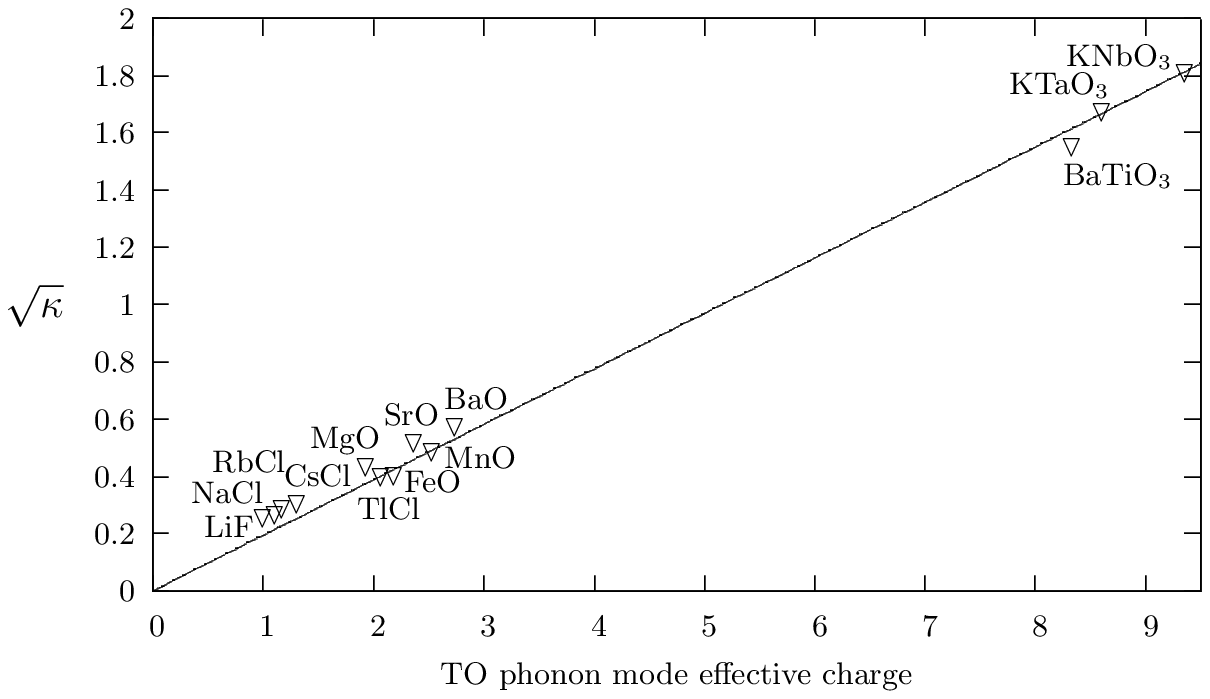}
\end{center}
\end{figure*}

\end{document}